# Application of new probabilistic graphical models in the genetic regulatory networks studies

Junbai Wang[1][*], Leo Wang-Kit Cheung[2] and Jan Delabie[3]

1. Department of Biological Sciences, Columbia University, 1212 Amsterdam Avenue, MC 2442, New York, NY 10027, USA

2. Epidemiology Section, Cancer Etiology Program, Cancer Research Center of Hawaii, University of Hawaii, 1236 Lauhala Street, Honolulu, HI 96813, USA

3. Department of Pathology, The Norwegian Radium Hospital, N0310 OSLO, Norway

Junbai Wang (**jw2256@columbia.edu**)

Leo Wang-Kit Cheung (**lcheung@crch.hawaii.edu**)

Jan Delabie (**jan.delabie@labmed.uio.no**)

**\* Corresponding author**

# Abstract

This paper introduces two new probabilistic graphical models for reconstruction of genetic regulatory networks using DNA microarray data. One is an Independence Graph (IG) model with either a forward or a backward search algorithm and the other one is a Gaussian Network (GN) model with a novel greedy search method. The performances of both models were evaluated on four MAPK pathways in yeast and three simulated data sets. Generally, an IG model provides a sparse graph but a GN model produces a dense graph where more information about gene-gene interactions is preserved. Additionally, we found two key limitations in the prediction of genetic regulatory networks using DNA microarray data, the first is the sufficiency of sample size and the second is the complexity of network structures may not be captured without additional data at the protein level. Those limitations are present in all prediction methods which used only DNA microarray data.

# Introduction

DNA microarray technology provides the expression levels of thousands of genes simultaneously. These expression data are static and hence do not give insight of how genes interact with each other. Therefore, it is a challenge to extract valuable gene-gene interaction information from such a large amount of microarray expression data. Mathematical and computational modeling is becoming increasingly important as a tool to capture gene interactions from expression data [1]. This information can be used as a basis for treating and diagnosing diseases. It may also contribute substantially to our basic understanding of biological processes [2, 3].

Recently, many mathematical and computational approaches for modeling gene regulations have been proposed, such as Boolean networks [4], Bayesian networks [5], the S-system [6], the Gaussian graphical model [7], and models based on support vector machines [8] and partial or ordinary differential equations et al, [1]. Boolean networks and support vector machines assume that genes are either ON or OFF (i.e. in binary expression values), whereas Bayesian networks are often based on more general discrete expression values [5]. Nevertheless, the examination of real gene expression measurements shows that gene expression levels tend to be continuous rather than in binary or discrete values. Some information may be lost if one employs models that do not take advantage of the continuous nature of expression measurements. The S-system and the Gaussian graphical model have been applied to continuous DNA micorarray expression data with limited success, i.e. the S-system is well suited for the time series data but does not handle the human tissue samples, in addition, it requires a huge amount of computational resources. The Gaussian graphical model is very sensitive to the rank order of input matrix and does not provide information about the causal relation between paired genes, which is essential for the reconstruction of genetics networks.

To overcome the shortcomings of above-mentioned methods, we have developed two new models for the prediction of genetic regulatory networks using DNA microarray data, a new Independence graph (IG) model and a new Gaussian network (GN) model. Our proposed models are based on probability theory and graph theory, which deal with uncertainty and complexity that are inherent in microarray experiments. The new IG model is based on an extension of the undirected Gaussian graphical model [9] for network structure learning, it performs searching for a directed acyclic graph (DAG) inside an undirected graph by using some orientation rules of graph [10]. Our GN model [11] combines a scoring metric with a novel search procedure. The scoring metric takes a network structure, microarray expression profiles, and also can be used in combination with user's prior knowledge of networks (e.g. the molecular pathways in our study). The model then returns a score which is proportional to the posterior probability of the network structure given the expression data. Then, the search procedure generates other network structures for evaluation by the scoring metric [11].

Conducting DNA microarray experiments on a series of time points following a physiological event provides us time series data to examine temporal changes in gene expression. The interest is to study the effect of time that it needs for the regulatory gene to express its protein product and the transcription of the target gene to be affected (directly or indirectly) by this regulator protein. As a result, we may observe a statistically significant correlation between the expression of a regulator and its target if biologically relevant time slices are used [12, 13, 14]. Thus, we tested our proposed models on three simulated data sets and several known gene regulatory networks, i.e. four MAPK pathways, by using time-series DNA microarray expression profiles. The aim of this work is to evaluate the robustness of our newly developed models.

## Methods

### Independence Graphs

Independence graphs (IG) are the probability models for multivariate random observations whose independence is characterized by a graph, G=(V, E), where V is a finite set of vertices and E is a finite set of edges. The independence graph is defined by pair-wise Markov properties, where there is no edge between two vertices whenever the pair of variables is independent given all the remaining variables. The resulting undirected independence graph gives a picture of the pattern of dependence or association between the variables [15].

### Independence graph with a forward search algorithm

Given an independence graph, G, and a k-dimensional continuous random vector X, with a multivariate normal distribution, we then use a covariance selection model to search for the best independence graph consistent with the data. The conditional independence constraints are equivalent to specifying zeros in the parameters in the inverse of the covariance matrix corresponding to the absence of an edge in G [7]. In other words, two variables are independent given remaining variables if and only if the corresponding element of the inverse of the covariance matrix is zero [15].

The independence graph with a forward search algorithm (IGF) can be used to analyze microarray gene expression data in the follow steps:

1. $X=(X_1, X_2, \ldots, X_k)$ is a k-dimensional vector, k is the number of genes. An initial graph G is built, i.e. an empty graph G with k vertices corresponds to k genes.

2. An iterative algorithm [9] for computing maximum likelihood estimates of the covariance matrix, Cov(G), of the initial graph G is then applied.

3. An edge $E_i$ is added into the initial graph, and a new covariance matrix, denoted as $Cov(E_i)$, is estimated by the iterative maximum likelihood estimates. Then, the significance of the added edge is tested by the deviance difference ($dev_i=N*[\log(\det(cov(G)))- \log(\det(cov(E_i)))]$, where N is the number of samples and i represents the i-th possible pair-wise edge of G). The deviance difference $dev_i$ has an asymptotic Chi-square distribution with one degree of freedom.

4. The most significant probability value is selected from $dev_i$. If it is smaller than the predefined significant threshold (e.g. significance level P = 0.05) then the corresponding edge is added to the initial graph G and step 2 is reiterated. If the probability value is larger than the threshold, the search is terminated and then the current undirected independence graph is retained.

5. Orientation rules [10] were used to find a directed acyclic graph (DAG) from the resulting undirected independence graph from step 4.

6. The final result is a DAG, where vertices represent genes, edges depict associations between a pair of genes, and the arrows explain the cause and effect between a pair of genes. Hence, the DAG may reveal the genetic regulatory networks.

NOTE: Cov(G) represents the covariance matrix of the current graph G, and Cov($E_i$) represents the estimated covariance matrix after one edge $E_i$ is added.

By starting from an empty initial graph, the IGF algorithm is not so sensitive to the rank order of input matrix. Therefore, we may still apply the IGF to predict gene-gene interactions when the number of sample size is fewer than the number of genes.

**Independence graph with a forward depth-limited search algorithm**

The IGF algorithm applies an exact search method that calculating the maximum likelihood estimates of the covariance matrix at both steps 2 and 3 during each iteration. It requires a considerable amount of computations when the number of genes is large. To avoid such heavy calculations, we also developed a relatively fast algorithm by combining the independence graph with an approximate search algorithm, that we called a forward depth-limited search algorithm (IGFD).

Below is a description of the IGFD strategy:

1. $X=(X_1, X_2, \ldots, X_k)$ is a k-dimensional vector, where k represents the number of genes. We built an initial graph, i.e. an empty graph G with k vertices corresponds to k genes.

2. An iterative algorithm [9] is then applied for computing the maximum likelihood estimates of the covariance matrix Cov(G) of the initial graph. A new covariance matrix Cov($E_i$) is then estimated after one edge $E_i$ has been added into the graph G. Subsequently, the deviance difference is used to measure the significance of the added edge ($dev_i = N*[\log(\det(cov(G))) - \log(\det(cov(E_i)))]$, where N is the number of samples and i represents all possible pair-wise edges). The deviance difference $dev_i$ has an asymptotic Chi-square distribution with one degree of freedom. Finally, all edges are sorted in descending order of their deviance differences and the label 0 is assigned for all sorted-edges.

3. The depth-limited search edges function is used to find out all possible edges that can be added into graph G with certain conditions, i.e. the search procedure is stopped after a manageable number of iterations or the most significant probability value of the remaining edges is above the threshold (i.e. P>0.7). This results into an undirected independence graph.

4. The depth-limited search edges function: Input (sorted-edges and graph G), Output (updated sorted-edges and updated graph G).

    (1) M = length of sorted-edges.

    (2) For i=1 to M

        If the label of sorted-edges(i) is 0

            Add sorted-edges(i) to graph G then use the deviance difference to measure the significance of the added edge.

                If the probability value < threshold (i.e. significance level P = 0.05) then remove sorted-edges(i) from sorted-edges, otherwise assign label 1 to sorted-edges(i) and remove this edge from G.

        End if

    End for

(3) If all labels of sorted-edges are 1 then assign 0 to them.

(4) Return sorted-edges and G.

5. Orientation rules [10] are used to find a directed acyclic graph (DAG) from the undirected independence graph.

NOTE: Cov(G) represents the covariance matrix of the current graph G, and Cov($E_i$) represents the estimated covariance matrix after one edge $E_i$ is added.

The differences between the IGF and the IGFD are their model search methods, where the IGF used a step by step search procedure but the IGFD applied an approximated search strategy with a specified significant probability threshold. In the IGFD, the calculation of the maximum likelihood estimates of the covariance matrix for all possible edges is only required at step 2 which significantly speed-up the whole search procedure (Tables 1, 2, 3, 4). For this reason, we may gain some special advantages when we apply the IGFD to learn a large genetic regulator network using gene expression data.

**Independence graph with a backward search algorithm**

The backward search algorithm (IGB) of the conditional independence graph has been studied by a number of authors before, its detailed description is described elsewhere [7, 9]. In this work, we expanded the resulting undirected graph to a directed acyclic graph (DAG) by implementing the orientation rules [10], where DAG may assist us to explore the cause and effect between a pair of genes.

**Orientation rules of graph**

Try to find a graph $G_1$, which is a consistent DAG extension of the undirected graph G:

1. Input the undirected graph G, where V is a set of vertices in G, and A, B, S are disjoint subsets of V. If there exists a set S⊆V\{A, B} such that A ⊥ B | S (A is independent of B given S), then let Sep(A, B)=S same as removes the edge between A and B. Here, a Fisher's Z-transformed test [16] is used to test if the correlation is statistically significant and check the conditional independence i.e. A ⊥B | S.

2. For all unshielded triples <A, B, C> (A is adjacent to B, B is adjacent to C, and A is not adjacent to C) in G, follow the orientation rule R0 (Figure 1) to orient A —> B and C —> B if B ∉ Sep(A, C) or in other words if and only if A, C are dependent conditional on every set containing B but not A, C. For more details, please refer to a similar application of this rule in the SGS algorithm [17].

3. Find a partially directed graph $G_1$ by using the other 4 orientation rules R1, R2, R3, R4 (Figure 1).

4. Find a consistent DAG extension of $G_1$ in the following steps

    a. If $G_1$ has no un-oriented edges then STOP

    b. Choose an un-oriented edge A — B from $G_1$

    c. Orient A —> B in $G_1$ and close orientations under rules R1, R2, R3, R4

    d. Go to step a

Detailed description and proof of the orientation rules are available in reference [10].

The original idea of our IG algorithms (independence graph) came from the early work by Verma and Pearl [18], where they constructed an undirected independence graph before searching for a directed acyclic graph. Later, Spirtes et al. proposed a variation of their idea [17] by setting the initial graph in their PC algorithm to the undirected independence graph, rather than the complete undirected graph and then proceed in the same way. They called this algorithm IG (independence graph). In our work, we used the same strategy as they did, but we applied alternative methods to build the undirected independence graph, i.e. a Gaussian graphical model, and a search for the directed acyclic graph with Meek's origination rules [10, 19]. To our knowledge, this work is the first in the literature to have a Gaussian graphical model combined with graph orientation rules. In addition, we will explain shortly about the Chi-square test that is used in the IG algorithms, where the model selection procedure is based on the deviance difference between the initial model $M_0$ and the estimated model $M_i$ (i.e. having added one edge $E_i$ into the initial graph G). In other words, it is the difference between the maximized log likelihood value under model $M_0$ and the maximized log likelihood value under $M_i$. Thus, the deviance difference is $dev_i = N*[\log(\det(cov(G))) - \log(\det(cov(E_i)))]$ which have been used by a number of authors before [7, 9]. Under $M_0$, $dev_i$ has an asymptotic Chi-square distribution with degree of freedom given as the difference in number of free parameters between $M_0$ and $M_i$. Its derivation can be found in Edwards's book [15].

**Gaussian networks**

We consider $\mathbf{X}=(X_1, \ldots, X_n)$ to be a set of random variables (genes), where $x_i$ denote a value of $X_i$, the i-th component of $\mathbf{X}$. A probabilistic graphical model for $\mathbf{X}$ is a graphical factorization of the joint probability density function,

$$p(X_1, \ldots, X_n) = \Pi_{(i=1\ldots n)}\, p(X_i \mid \pi_i, \theta_i)$$

For every variable $X_i$, $\pi_i$ represents the parents of $X_i$, and $\theta_i$ represents a finite set of local parameters, $\theta_i = (m_j, b_{ji}, v_i)$, where $m_j$ is the unconditional mean of $X_j$, $b_{ji}$ is the linear coefficient reflecting the strength of the relationship between $X_i$ and $X_j$, and $v_i$ is the conditional variance of $X_i$ given values for $\pi_i$ [11]. The network structure $B_S$ for $\mathbf{X}$ is a directed acyclic graph (DAG), which indicates the assertions that $X_i$ and $\{X_1, \ldots, X_n\} \setminus \pi_i$ are independent given $\pi_i$. Here, we assume that each variable is continuous, and each local density function is a linear regression model:

$$p(X_i \mid \pi_i, \theta_i) = N(X_i;\, m_i + \Sigma_{(X_j \in \pi_i)}\, b_{ji}(X_j - m_j),\, 1/v_i).$$

Given this form, a missing arc from $X_j$ to $X_i$ implies that $b_{ji}=0$ in the linear-regression model. And the resulting probabilistic graphical model is a Gaussian network (GN).

Geiger and Heckerman had described a scoring metric for Gaussian networks with continuous variables. The metric is based on the fact that the normal-Wishart distribution is conjugate with respect to the multivariate normal distribution. This allows us to obtain a closed formula for the computation of the marginal likelihood of the data given the structure. The detailed description of the scoring metric and the validity is previously described [11]. From this scoring metric, it can be proved that the marginal likelihood for a general Gaussian network can be calculated by:

$$p(D \mid B_S) = \Pi_{(i=1\ldots n)} \, p(D^{[X_i, \pi_i]} \mid B_{Sc}) / p(D^{\pi_i} \mid B_{Sc})$$

Where each term is of the form,

$$p(D \mid B_{Sc}) = (2\pi)^{-nN/2} \, (\nu/(\nu+N))^{n/2} \, C(n,\alpha)/C(n, \alpha+N) \, |T_0|^{\alpha/2} \, |T_N|^{-(\alpha+N)/2}$$

$D^{[X_i, \pi_i]}$ is the dataset (all instances of **X**) D restricted to the variables $X_i \cup \pi_i$ and $C(n,\alpha)$ defined as follow:

$$C(n,\alpha) = [\, 2^{\alpha n/2} \, \pi^{n(n-1)/4} \, \Pi_{(i=1\ldots n)} \, \Gamma((\alpha+1-i)/2) \,]^{-1}$$

Thus, we obtain a metric for scoring the marginal likelihood of an arbitrary Gaussian network structure $B_S$. The discussion of relevant parameters $\alpha$, $\nu$, $T_0$, $\mu_0$ and the prior probabilities $P(B_S)$ for learning Gaussian networks is shown in [11]. For our purpose, we assume the prior probabilities of all network structures are equally likely.

Since the number of Gaussian network structures grows very rapidly as a function of the variables [21], we implemented a novel search method -- a combination of partial correlation coefficients [22] with an iterative greedy hill-climbing algorithm -- to find a good solution in large spaces. For this search method, in each step we apply a greedy search algorithm [20] until it reaches a local maximum, then we perturbs the current network structure according to the significance of a pair of edges and repeat the search for a manageable number of iterations, i.e. change the structure of a pair of edges if its partial correlation coefficients belongs to L percentages of the least significant edges, L is the learning rate which defined as a function $L=\alpha(\beta/\alpha)^{i/n}$ where $\alpha$ is the maximum perturbation rate, $\beta$ is the minimum perturbation rate, i=0,1,2…n and n is the number of iterations (α=0.8, β=0.2 and n=20 in this study). This search method avoids the greedy search algorithm getting stuck at a local maximum and often provides better results than the random perturbation of network structure [20].

Finally, we would like to clarify some potential misunderstandings between the IG algorithm and the Bayesian search approach, i.e. the Gaussian networks. The IG approach [17, 18] uses tests of conditional independence to construct sets of DAGs that impose the same conditional independence relations. The Bayesian search approach [5, 11] uses a Bayesian scoring metric combined with a search algorithm to look for the DAG with the highest posterior probability. Hence, the IG approach and the Bayesian search approach are two distinct approaches to learn the DAGs [20].

**Evaluation methodology**

The methodology [20] for measuring the learning accuracy of various algorithms is as follows: we quantify the learning accuracy by measuring the difference of network structures between the true networks (i.e. the expert's domain knowledge of the pathway) [23] and the predicted networks. The results are scored by three measures: (1) the percentage of edge existence errors of commission (i.e. the number of edges that are adjacent or connected in the predicted network but not in the true network divided by the total number of edges in the true network), (2) the percentage of edge direction errors of commission (i.e. the number of edges that have different arrow directions between the predicted network and the true network divided by the total number of edges in the true network), (3) the percentage of edge existence errors of omission (i.e. the number of edges that are adjacent or connected in the true network but not in the predicted network divided by the total number of edges in the true network).

Some edge existence errors of commission are more informative than others [24]. They link genes that are not in a direct parent-child relationship but are still nearby in the pathway, and those informative edges are going to provide crucial information when we explore some unknown biological pathways. For this reason, the informative edges will not be considered in our edge existence errors of commission (i.e. predicted pathways A->C, A->D or B->D will not be included in edge existence errors of commission if the true pathway is A->B->C->D).

## Results using real experimental data

**Objective:** We applied an independence graph model with a forward search algorithm (IGF), an independence graph model with a forward depth-limited search algorithm (IGFD), an independence graph model with a backward search algorithm (IGB), and a Gaussian networks (GN) model with a novel greedy search method on microarray gene expression data to explore the potential gene-gene interactions. All models are implemented in MATLAB, and run under Windows 2000 operating system of a portable PC with one Pentium processor.

**Sources of experimental data for the MAPK pathways:** The microarray data for the Mitogen-activated protein kinase (MAPK) pathways in yeast were obtained from the web supplement of the publication of Roberts et al. [23]

(http://www.rii.com/publications/2000/s287873.htm). This data set has a large number of time points (46 time points) and had been studied by a number of authors before [25, 26]. In addition, the MAPK pathways are among the most thoroughly studied networks in yeast, thus making them perfect for testing our newly developed approaches and for verifying the biological relevance of the recovered networks.

**Processing of data for the MAPK pathways:** 46 experiments and 6221 genes were considered in this study. There were less than 20% missing values across all experiments. The preprocessing of microarray data was the same as in the original publication. The missing data were imputed by the K-nearest neighbour method [27] and the raw ratios were log10 transformed before further analysis. From this dataset, we selected 13 genes (WSC1, WSC2, WSC3, MID2, RHO1, PKC1, BCK1, MKK1, MKK2, MPK1, SWI4, SWI6 and RLM1) to reconstruct the PKC pathway, 11 genes (SHO1, RAS2, CDC42, STE20, STE11, STE7, KSS1, RST1, RST2, STE12 and TEC1) to reconstruct the Filamentous pathway, 13 genes (SLN1, SHO1, YPD1, SSK1, STE20, STE11, SSK2, SSK22, PBS2, HOG1, MSN2, MSN4 and MCM1) to learn the HOG pathway, and 15 genes (STE2, STE3, STE4, STE18, STE5, CDC42, STE20, BNI1, STE11, STE7, FUS3, FAR1, RST1, RST2 and STE12) to reconstruct the Pheromone pathway. The gold standards of the four MAPK pathways were adopted from the original publication [23] (Figure 2).

**Model evaluation and prediction of the MAPK pathways:** From the result of structure learning of the four MAPK pathways, we found that the IGFD is the most efficient model. It provided the same results as the IGF, but required much less learning time, i.e. 4 to 22 seconds (Tables 1, 2, 3, 4). On the other hand, the GN model gave the most accurate results for all tested pathways (Tables 1, 2, 3, 4), but it required the longest learning time, i.e. 494 to 2130 seconds for no predefined gene ordering (Tables 1, 2, 3, 4). The true network structures of the PKC, the Filamentous growth and the HOG pathway are rather similar (Figure 2), except that there are negative gene regulations in the Filamentous growth and the HOG pathways. This may explain the high percentages of EEEO in the Filamentous growth (29.4% to 58.8%) and the HOG (23.8% to 57.1%) pathways (Tables 2, 3), but relatively low percentages of EEEO (3.7% to 25.9%) in the PKC pathway (Table 1). For the percentages of EEEC, there are no big differences among the PKC (3.7% to 11.1%), the Filamentous growth (5.9%) and the HOG (9.5% to 14.3%) pathways (Tables 1, 2, 3) across all models. At the end, we found that the Pheromone pathway has the most complicated network structure (Figure 2), which had a surprisingly high percentage of EEEO (44% to 72%) and its percentage of EEEC was nearly two times higher than the rest of the MAPK pathways (Table 3). Besides, we had gained only a little improvement in its network structure learning when we applied a less stringent significance level, i.e. $P = 0.3$., (Table 4) in the IGFD.

Finally, we applied existing models such as the Boolean Network (BN), the Dynamic Bayesian Network (DBN) and the Linear Differential Model (LDM) [28] to the same datasets, i.e. the microarray data of four MAPK pathways that had been analyzed by our models. The identical evaluation methodology [20] had also been applied in order to illustrate the accuracy of their predictions (Tables 1, 2, 3, 4). We found that all methods had a similar trend in their error rates. That is according to the complexity of the true pathway structures, the EEEC of the DBN and the LDM had the same structure throughout all pathways which were approaching the best results that had been obtained by our GN and IGFD models. However, the BN had extremely high EEEC for the last two pathways, and the EEEO of the DBN and the LDM were nearly two times higher than the GN and the IGFD. Generally, the BN has the poorest performance among all models. This had also been noted in early studies [28]. The DBN and the LDM often retrieve much fewer gene-gene interactions than the GN and the IGFD models (Tables 1, 2, 3, 4). Overall, our newly developed models (GN and IGFD) outperformed those previously existing models in this study. Moreover, we noticed that the network structure of various pathways potentially plays an important role in the reconstruction of genetic regulatory networks using DNA microaray data. The complexity of genetic regulatory systems strongly influences the accuracy of our predictions, regardless of what kinds of learning algorithms are used.

## Results using simulated data

**Objective:** we used IGFD, the most efficient model of current proposed approaches, and the well know PC algorithm [17] on three simulated data sets to evaluate the accuracy of our newly developed method.

**Source of simulated data:** Based on the structure of three randomly produced directed acyclic graphs (DAGs), we generated the sample data by using the TETRAD program [36]. The three DAGs involve eight vertices, and a completely different network structure and the size of sample data i.e. 40, 200 and 1000 samples at here.

**Model evaluation and prediction of randomly generated DAG:** We compared the performance of IGFD with the PC algorithm [17] by applying the same model evaluation method [20] as used in the early section, where the gold standard comes from the topology of three DAGs that produced sample data for this study. The result in (Figure 3) shows that the edge existence errors of commission (EEEC) and the edge direction errors of commission (EDEC) are not so sensitive to the number of samples that used in the learning. But the edge existence errors of omission (EEEO) are extremely dependent on the size of sample data. For example, both PC and IGFD achieved a nearly four fold improvement of their performance when the number of samples increased from 40 to 1000. This may also explains the reason of high EEEO in the prediction of genetic regulatory networks using real micorarray data (Table 1, 2, 3, 4). Since the number of available microarray experiments is not so adequate for all learning models, i.e. we only have 46 microarray experiments to learn the causal relations of 10 to 15 genes during this study. Overall, our IGFD model clearly outperformed the popular PC algorithm in (figure 3), where it has a constant lower EEEO and EDEC than the PC algorithm. We also found another key limitation in the reconstruction of gene regulatory networks using microarray data, the sufficiency of sample size, which may strongly affect the accuracy of our predictions.

## Conclusions and discussions

In this work, we have presented two novel models with completely different learning stratagems for the reverse engineering of gene regulatory networks using DNA microarray data: one is an independence graph (IG) with either a forward or a backward search algorithm (IGF, IGFD and IGB), and the other one is a Gaussian network (GN) model with a novel greedy search method. We had tested them on several known biological pathways, i.e. four MAPK pathways (Figure 2), and three simulated data sets (Figure 3). The GN model often provides the best prediction results (EEEC = 2.1% to 12%) and retrieves more gene-gene interactions than the IG models, (Tables 1, 2, 3, 4), but it suffers from a long learning time. For the IG models, with a stringent significance threshold (i.e. P=0.05), often give sparse graphs where a few of weak gene-gene interactions may be missed, this may explain the percentages of EEEO for IGFD (i.e., 18.5%, 35.3%, 57.1%, 56%) in (Tables 1, 2, 3, 4). However, we may obtain a result as good as the GN model if the significance level of the IG models is well selected, e.g. the EEEO of the EG+GN (3.7%, 35.3%, 23.8%, 48%) versus the EEEO (with P = or > 0.2) of the IGFD (11.1%, 29.4%, 28.6%, 52%) in (Tables 1, 2, 3, 4). On the other hand, the IG models may have fewer errors in retrieving gene-gene interactions i.e. the percentages of EEEC (with P=0.05) for the IGFD (7.4%, 5.9%, 9.5%, 12%) in (Tables 1, 2, 3, 4). Thus, the IG models probably have a particularly advantage for the exploration of unknown genetic regulatory systems using DNA microarray data. These models can be used to help formulate hypotheses in a probabilistic framework and allow their full implications to be investigated and recognized. Consequently, the wet-laboratory work and the clinical experimentations will then be concentrated at the validation of a small number of carefully chosen model predictions of crucial genes that may determine the genetic regulations encapsulated in the pathway of interest.

Additionally, we had compared performance of our newly developed models with previously existing models, i.e. the BN, the DBN, the LDM and the PC algorithm, where our models showed better prediction results. From (Figure 3), we knew that the accuracy of reconstructed networks is strongly correlated with the number of samples that are used in the learning. On the other words, the sample size will be a key factor to minimize the error rate of reconstructing gene regulatory networks using microarray data. We also carried out a systemic study on the performance of all suggested models against the complexity of true network structures. The result is illustrated in (Tables 1, 2, 3, 4), where all models have shown a clear trend of increasing their error rates as the complexity of true network structures is increasing. In other words, predicting a biological pathway with a simpler structure results in fewer errors (i.e. the PKC pathway), but predicting a pathway with a more complex structure results in more errors (i.e. the Pheromone pathway) (Tables 1, 4). Potentially, there are two reasons that may contribute to such trend: the sufficiency of mathematical modeling and the sufficiency of biological measurements. In the first place, our proposed models are not able to distinguish between a positive gene regulation from a negative gene regulation, and do not handle the gene regulation with a feed-backward loop. However, it is well known that gene-gene interactions with feed-backward loops often exist in biological pathways or genetic regulatory systems. Thus, the insufficiency of present mathematical modeling (IG and GN) may lead to an unsurprisingly high EEEO in these complex cases (Tables 1, 2, 3, 4). For instance, the PKC pathway has a similar network structure as the Filamentous Growth pathway (Figure 2), where the latter one has negative gene regulations, and a much higher EEEO for the Filamentous Growth pathway is obtained (Tables 1, 2). Secondly, for any real biochemical networks, there are three levels of observations (gene expression at the mRNA level, the protein level, and the metabolite level) will be needed for accurate descriptions of cellular biochemical system. It has been noted that genes often do not interact directly with other genes in any global biochemical network, but gene induction or repression often occurs through the actions of specific proteins. Gene expression can also be affected directly by metabolites or protein-metabolite complexes. In this work, we simplified the global biochemical network to a gene network (genetic regulatory or expression network), and assumed that the gene expression profiles of the regulator (i.e. a signaling molecule or a transcription factor) provide information on its activity levels. Such concordant changes in the expression of both the regulator and its targets might allow our approach to detect statistical associations between them [29]. Despite the fact that some important information can be obtained from mRNA expression data, the use of additional data at the protein level and the metabolite level with the mRNA expression data will be highly recommended when the network structure becomes more complex. We may then be able to achieve a more accurate view of a complex gene regulatory system when we have sufficient experimental measurements.

In conclusion, all of our proposed models (IG and GN) had shown promising results on three simulated data sets and the reconstruction of the MAPK pathways in yeast. We found two important factors, the sufficiency of sample size and the complexity of real network structure, which may highly influence the quality of reverse engineering of gene regulatory networks. In particular, the IGFD model not only has the algorithmic efficiency for network structure learning, but also has the flexibility to extract the network structure with various significance levels. The IGFD may become an attractive model for the prediction of large genetic regulatory systems using DNA microarray data.

Overall, the present research studies aimed to investigate whether our new models are applicable to experimental data in order to reconstruct genetic regulatory networks [14, 20, 29, 30, 31, 32]. We plan to keep refining our probabilistic models for network structure learning with the increase in sufficiency of biological measurements, i.e. combining appropriate experimental designs with the use of DNA microarrays, and integrating the data of binding sites of transcription factors with appropriately analyzed expression data and with data on the occurrence of upstream sequence motifs or protein-protein interactions [25, 30, 33, 34, 35].

## Acknowledgements

We would like to thank Ola myklebost for very helpful discussions and Marinus F. van Batenburg for critical reading of manuscript.

**Figure Legends**

**Figure 1 Simplified schematics of Orientation rules.** Orientation rules R0, R1, R2, R3, R4 that have been used in independence graph models to find a consistent DAG extension of undirected graph.

**Figure 2 Simplified schematics of four MAPK pathways in yeast.** Lines show direct effects, arrows imply positive regulation, and bars imply negative regulation.

**Figure 3 The percentages of the edge existence errors of commission, the edge existence errors of omission and the direction errors of commission in relation to the sample size of three randomly generated directed acyclic graphs.** PC: Spirte's PC algorithm; IGFD: our independence graph with forward depth-limited search algorithm; P denotes the significance level of a Chi-square test.

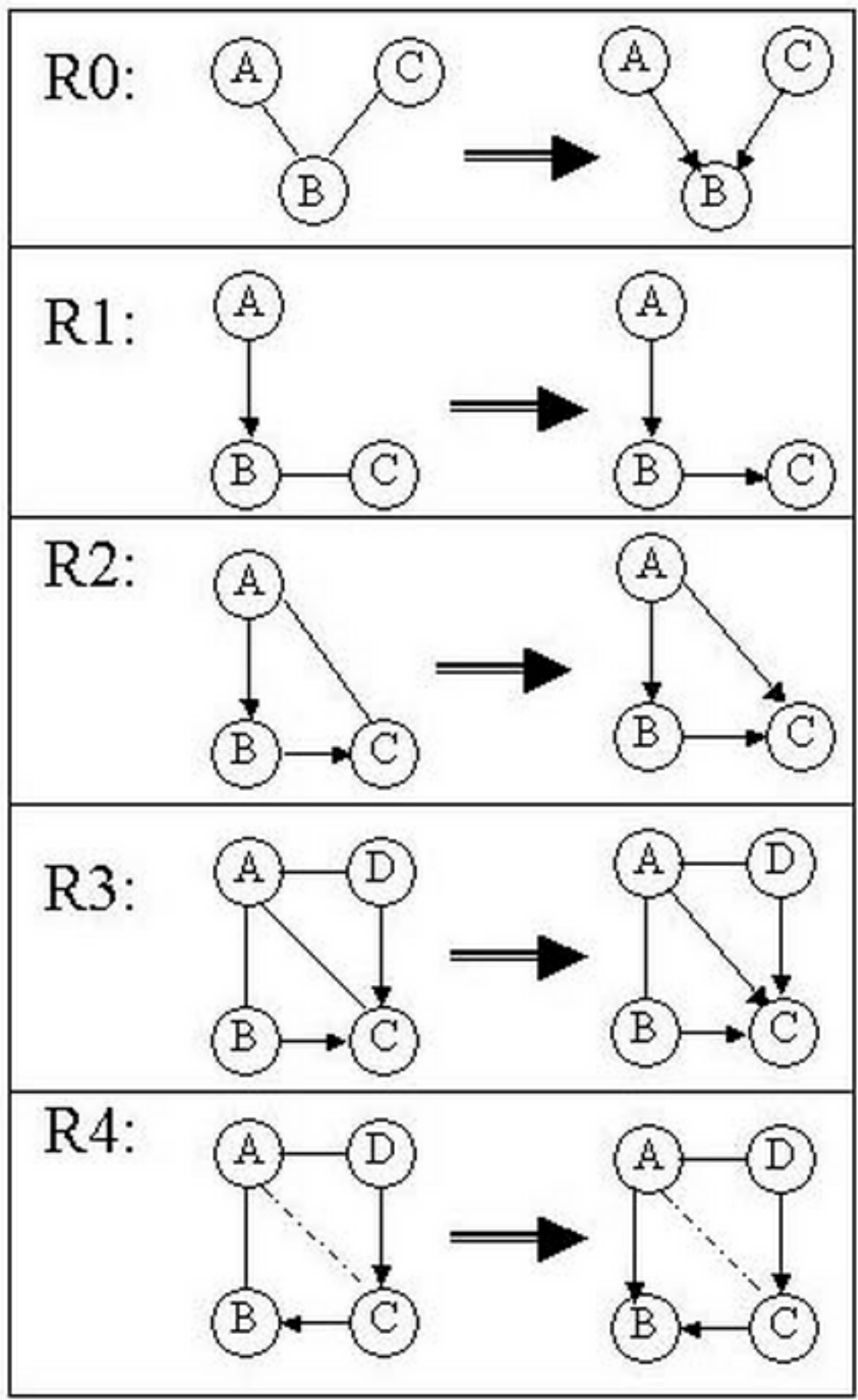

**Figure2**
[Click here to download high resolution image](#)

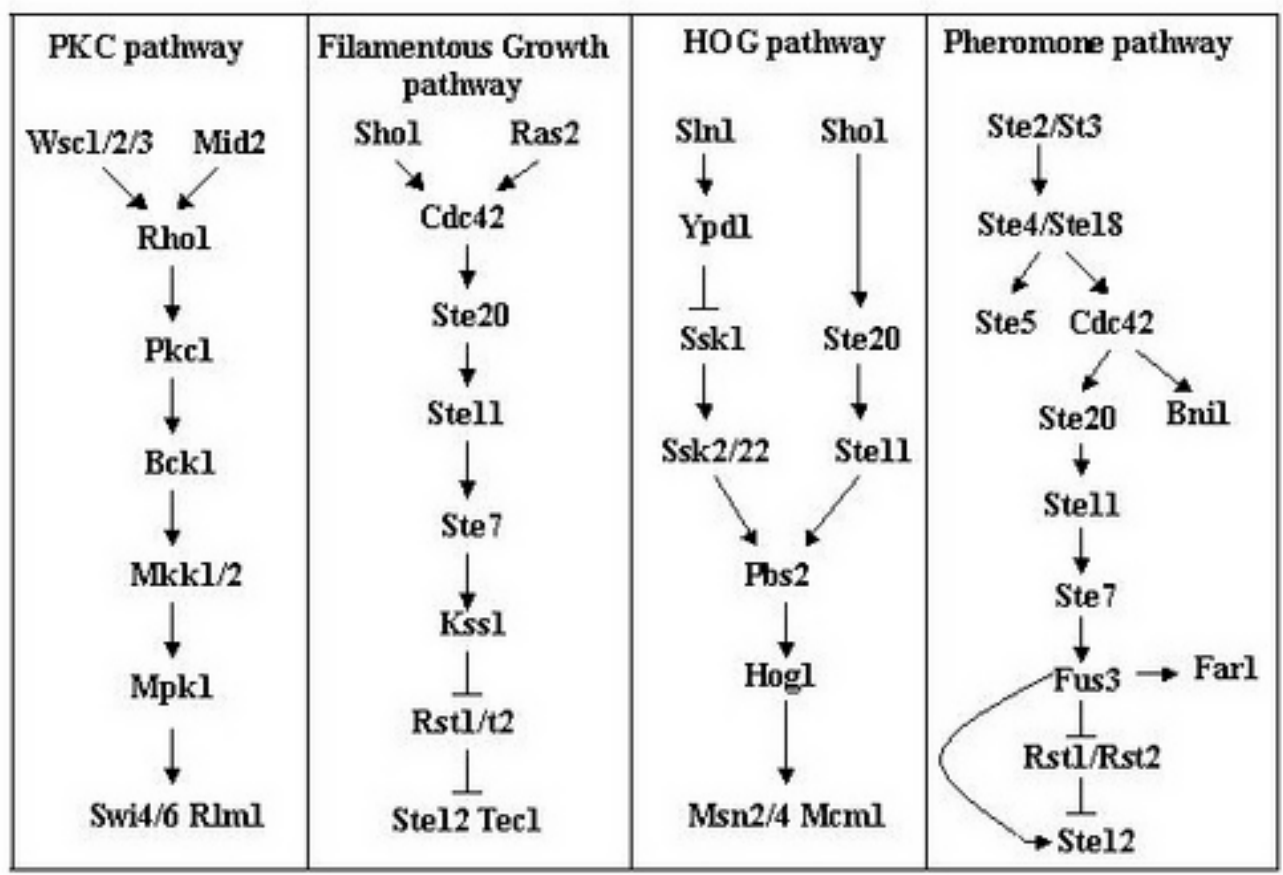

**Figure3**
Click here to download high resolution image

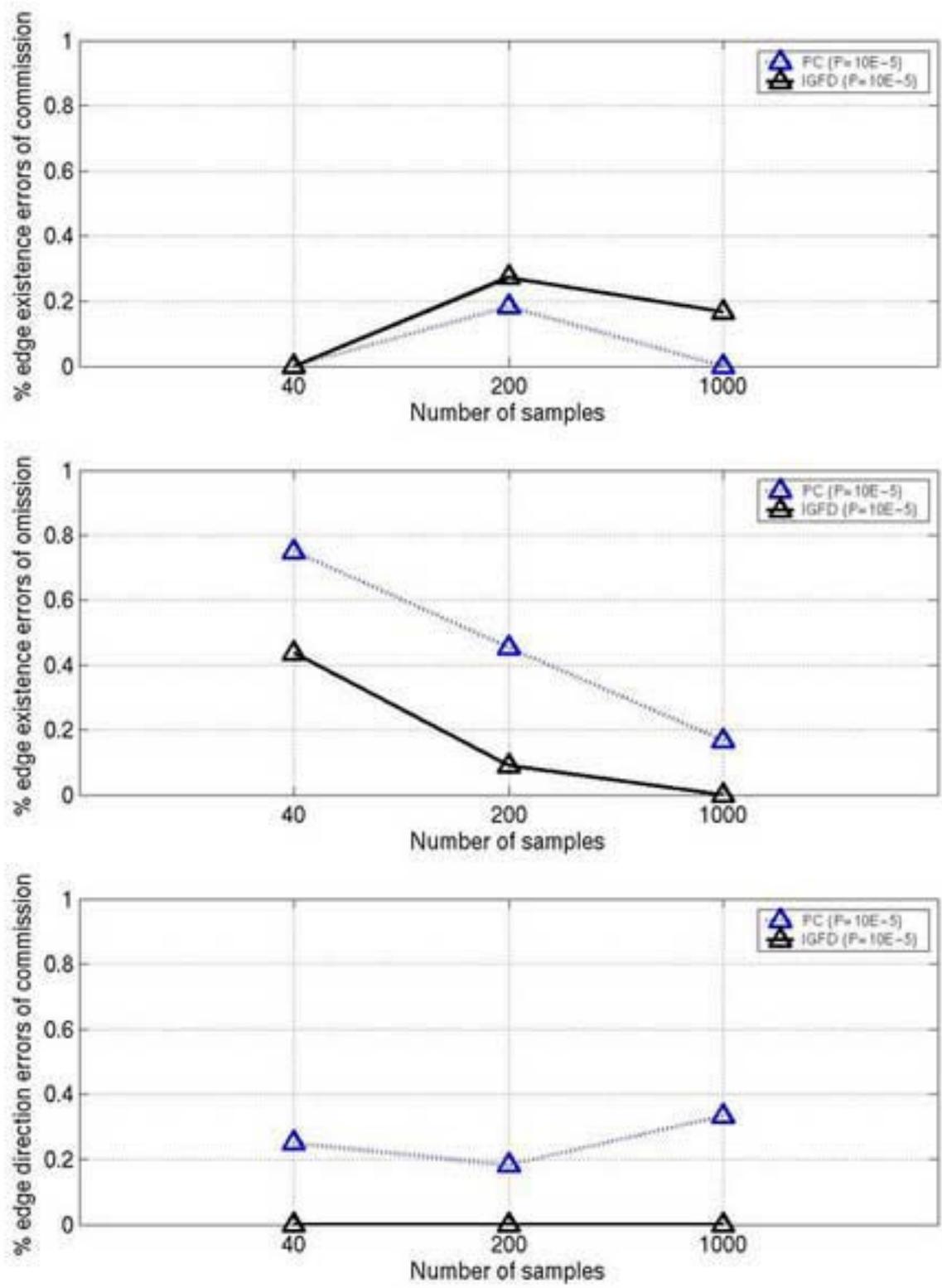

**Table**

| PKC pathway (13 genes, 46 time points) | %edge existence errors of commission | %edge direction errors of commission | %edge existence errors of omission | CPU time (second) |
|---|---|---|---|---|
| BN | 7.4 | 3.7 | 44.4 | NA |
| DBN | 0 | 0 | 48.2 | NA |
| LDM | 0 | 0 | 48.2 | NA |
| IGFD: (P=0.05, OV) | 7.4 | 0 | 18.5 | 22 |
| IGFD: (P=0.2, OV) | 11.1 | 0 | 11.1 | 241 |
| IGB:  (P=0.05, OV) | 3.7 | 0 | 25.9 | 65 |
| IGB:  (P=0.15, OV) | 11.11 | 0 | 11.11 | 125 |
| IGF:  (P=0.05, OV) | 7.4 | 0 | 18.5 | 296 |
| EG +GN:    (OV) | 11.1 | 0 | 18.5 | 289 |
| IGFD + GN: (OV) | 11.1 | 0 | 22.2 | 302 |
| IGFD + GN: (UV) | 7.4 | 18.5 | 7.4 | 1299 |
| EG +GN:    (UV) | 11.1 | 11.1 | 3.7 | 1286 |

**Table 1 Comparison of learning errors for the PKC pathway.** BN: Boolean network; DBN: Dynamic Bayesian Network; LDM: Linear Differential Model; IGFD: independence graph with forward depth-limited search algorithm; IGB: independence graph with backward search algorithm; IGF: independence graph with forward search algorithm; EG+GN: Gaussian networks with empty initial graph; IGFD+GN: Gaussian networks by using the learning results of IGFD as initial graph. OV means predefined gene ordering and UV represents no predefined gene ordering. NA means not available. P denotes the significance level of a Chi-square test.

| Filamentous pathway (11 genes, 46 time points) | %edge existence errors of commission | %edge direction errors of commission | %edge existence errors of omission | CPU time (second) |
|---|---|---|---|---|
| BN | 11.8 | 35.3 | 17.7 | NA |
| DBN | 5.9 | 0 | 70.6 | NA |
| LDM | 0 | 11.8 | 58.8 | NA |
| IGFD: (P=0.05, OV) | 5.9 | 0 | 35.3 | 7 |
| IGFD: (P=0.3, OV) | 5.9 | 0 | 29.4 | 49 |
| IGB:   (P=0.05, OV) | 5.9 | 0 | 52.9 | 16 |
| IGB:   (P=0.2, OV) | 5.9 | 0 | 41.2 | 18 |
| IGF:   (P=0.05, OV) | 5.9 | 0 | 35.3 | 24 |
| EG +GN:    (OV) | 5.9 | 0 | 58.8 | 86 |
| IGFD + GN: (OV) | 5.9 | 0 | 58.8 | 93 |
| IGFD + GN: (UV) | 5.9 | 17.7 | 35.3 | 494 |
| EG+GN:    (UV) | 5.9 | 5.9 | 35.3 | 504 |

**Table 2 Comparison of learning errors for the Filamentous pathway.** BN: Boolean network; DBN: Dynamic Bayesian Network; LDM: Linear Differential Model; IGFD: independence graph with forward depth-limited search algorithm; IGB: independence graph with backward search algorithm; IGF: independence graph with forward search algorithm; EG+GN: Gaussian networks with empty initial graph; IGFD+GN: Gaussian networks by using the learning results of IGFD as initial graph. OV means predefined gene ordering and UV represents no predefined gene ordering. NA means not available. P denotes the significance level of a Chi-square test.

| Hog pathway (13 genes, 46 time points) | %edge existence errors of commission | %edge direction errors of commission | %edge existence errors of omission | CPU time (second) |
|---|---|---|---|---|
| BN | 42.9 | 4.8 | 42.9 | NA |
| DBN | 4.8 | 0 | 61.9 | NA |
| LDM | 0 | 4.8 | 57.1 | NA |
| IGFD: (P=0.05, OV) | 9.5 | 0 | 57.1 | 4 |
| IGFD: (P=0.2, OV) | 14.3 | 0 | 28.6 | 44 |
| IGB:  (P=0.05, OV) | 14.3 | 0 | 47.6 | 27 |
| IGB:  (P=0.15, OV) | 14.3 | 0 | 33.3 | 33 |
| IGF:  (P=0.05, OV) | 9.5 | 0 | 57.1 | 19 |
| EG+GN:    (OV) | 9.5 | 0 | 52.4 | 223 |
| IGFD + GN: (OV) | 14.3 | 0 | 52.4 | 230 |
| IGFD + GN: (UV) | 14.3 | 9.5 | 28.6 | 1148 |
| EG +GN:    (UV) | 9.5 | 9.5 | 23.8 | 1097 |

**Table 3 Comparison of learning errors for the Hog pathway.** BN: Boolean network; DBN: Dynamic Bayesian Network; LDM: Linear Differential Model; IGFD: independence graph with forward depth-limited search algorithm; IGB: independence graph with backward search algorithm; IGF: independence graph with forward search algorithm; EG+GN: Gaussian networks with empty initial graph; IGFD+GN: Gaussian networks by using the learning results of IGFD as initial graph. OV means predefined gene ordering and UV represents no predefined gene ordering. NA means not available. P denotes the significance level of a Chi-square test.

| Pheromone pathway (15 genes, 46 time points) | %edge existence errors of commission | %edge direction errors of commission | %edge existence errors of omission | CPU time (second) |
|---|---|---|---|---|
| BN | 44 | 36 | 4 | NA |
| DBN | 4 | 0 | 76 | NA |
| LDM | 4 | 4 | 68 | NA |
| IGFD: (P=0.05, OV) | 12 | 0 | 56 | 22 |
| IGFD: (P=0.3, OV) | 28 | 0 | 52 | 215 |
| IGB:  (P=0.05, OV) | 32 | 0 | 60 | 113 |
| IGB:  (P=0.15, OV) | 36 | 0 | 56 | 160 |
| IGF:  (P=0.05, OV) | 12 | 0 | 60 | 250 |
| EG +GN:  (OV) | 4 | 0 | 68 | 352 |
| IGFD + GN: (OV) | 12 | 0 | 72 | 371 |
| IGFD + GN: (UV) | 20 | 8 | 44 | 2130 |
| EG +GN:  (UV) | 12 | 0 | 48 | 2130 |

**Table 4 Comparison of learning errors for the Pheromone pathway.** BN: Boolean network; DBN: Dynamic Bayesian Network; LDM: Linear Differential Model; IGFD: independence graph with forward depth-limited search algorithm; IGB: independence graph with backward search algorithm; IGF: independence graph with forward search algorithm; EG+GN: Gaussian networks with empty initial graph; IGFD+GN: Gaussian networks by using the learning results of IGFD as initial graph. OV means predefined gene ordering and UV represents no predefined gene ordering. NA means not available. P denotes the significance level of a Chi-square test.